# Geographic Space as Manifolds

Hezhishi Jiang[1], Liyan Xu[2]*, Tianshu Li[2], Jintong Tang[3], Zekun Chen[4], Yuxuan Wang[2], Hongmou Zhang[5]*, Yu Liu[3]*

    1 Academy for Advanced Interdisciplinary Studies, Peking University, Beijing, China

    2 College of Architecture and Landscape Architecture, Peking University, Beijing, China

    3 School of Earth and Space Sciences, Peking University, Beijing, China

    4 School of Mathematical Sciences, Peking University, Beijing, China

    5 School of Government, Peking University, Beijing, China

These authors contributed equally: Hezhishi Jiang, Liyan Xu, and Tianshu Li

These authors jointly supervised this work: Liyan Xu, Hongmou Zhang, and Yu Liu

# Abstract

The communications and interrelations between different locations on the Earth's surface have far-reaching implications for both social and natural systems[1–11]. Effective spatial analytics ideally require a *spatial representation*[12], where geographic principles are succinctly expressed within a defined metric space. However, common spatial representations, including map-based[13] or network-based[14] approaches, fall short by incompletely[14,15] or inaccurately[13] defining this metric space. Here we show, by introducing an inverse friction factor that captures the spatial constraints[14] in spatial networks, that a homogeneous, low-dimensional spatial representation—termed the Geographic Manifold—can be achieved. We illustrate the effectiveness of the Geographic Manifold in two classic scenarios of spatial analytics –location choice and propagation, where the otherwise complicated analyses are reduced to straightforward regular partitioning and concentric diffusing, respectively on the manifold with a high degree of accuracy. We further empirically explain and formally prove the general existence of the Geographic Manifold, which is grounded in the intrinsic Euclidean low-dimensional statistical physics properties of geographic phenomena. This work represents a step towards formalizing Tobler's famous First Law of Geography[16] from a geometric approach, where a regularized geospace thereby yielded is expected to contribute in learning abstract spatial structure representations for understanding and optimization purposes[17,18].

# Main Article

The communications and interrelations between different positions on the Earth's surface have wide-ranging implications for social and natural systems. Referred to in geography as "spatial interactions", they are the basis for modeling transport and logistics[1,8], the spread of infectious diseases[9,10], journey-to-crime[2,4], accessibility of employment[11], the movement[3] and speciation of organisms[5,6], and the connectivity and communication of ecosystems[7]. All of the above spatial analytics are fundamentally based on a particular metric space, or in general, *spatial representation*[12], where position, distance, and interaction intensity are defined.

A good spatial representation enhances analytical processes by succinctly expressing geographic principles, a purpose that existing approaches are inadequate to meet. The prevailing practice utilizes spatial representations based on projected maps, which, while convenient, assume that interactions are governed by distance, as exemplified by the Gravitational Law[19]. However, distance as a proxy for spatial interaction only performs well on an ideal "*featureless plain*", a two-dimensional, Euclidean space that disregards the heterogeneity of geographic features [13,20,21], such as population, demographics, resources, and infrastructure. Clearly, this idealized space does not reflect the complexities of the real world, leading to both theoretical and practical challenges[9,22]. Nevertheless, shedding light from the *effective distance* which has been shown to be a more accurate proxy for spatial interaction[9], an opportunity emerges if we construct a metric space upon effective distance where homogeneity can be achieved.

Alternatively, there is the network-based spatial representation that permits arbitrary distances and interactions between two positions[14]. The approach offers a wealth of tools for modeling spatial interactions[14]; while with a similar form of the effective distance, it shares the latter's incompleteness in forming a metric space. Nevertheless, the incorporation of the spatial constraint[14,23,24]–such as the distance-decaying possibility of forming edges in a spatial network – presents a potency to complete the spatial representation, as it establishes the statistical physics relationship between distance and interaction, echoing Tobler's famous First Law of Geography[25]. Furthermore, the low-dimensionality law uncovered by network scientists[26] provides an additional chance of succinctness, contributing to the formation of an ideal spatial representation.

In this paper, we demonstrate that a manifold-based metric space constitutes such a spatial representation as long as the First Law holds. Referring to this manifold the *Geographic Manifold*, we prove its general existence in the statistical physics sense, and show it ensures the necessary homogeneity and low-dimensionality that facilitate elegant spatial analytics.

## Defining the Geographic Manifold

The geometric definition of manifolds, considering the reality of spatial analytics, requires the local Euclidean and metric space properties, where the latter is mainly about symmetry and triangular inequality[27]. All these properties depend on the distance on the manifold. Hence, we first give the distance definition on the Geographic Manifold:

$$d_{ij} = f(\Phi_{ij}) = f\left(\frac{\lambda_{i,j}}{g(S_i, S_j)}\right) \quad (1)$$

where $i, j$ are two positions, or two nodes; $\lambda_{i,j}$ is the spatial interaction intensity; $S_i$ is the node "mass" characteristics such as population, economy, node degree, node strength, etc.; the function $g$ is monotonically increasing with respect to $S$ and is symmetric with respect to $i, j$, and usually takes the form of a power function; and the function $f$ is a distance function that is always greater than 0 when $i \neq j$ and is always equal to 0 when $i = j$. $\Phi_{ij}$ is the friction factor in spatial interaction, which includes the distance, etc. Note that the larger $\Phi_{ij}$ is, the larger the interaction intensity $\lambda_{i,j}$ will be, we call $\Phi_{ij}$ the *inverse friction factor*. Theoretically, $\Phi_{ij}$ can be regarded as the *noise* caused by "friction" in the generalized gravity model[28] in urban science, or as a generalization of the *normalized vectors* in the vector field theory[29,30] of human mobility. We argue that $\Phi_{ij}$ is the key to defining Geographic Manifolds, as: (1) Friction effects underlie the spatial constraints of spatial networks, and their general form is that as the distance increases, the cost or the resistance of forming edges increases [23,31,32]. Considering that extremely strong spatial constraints bring about planar graphs[31,32], which are clearly low-dimensional, and that $\Phi_{ij}$ filters the spatial constraints, $\Phi_{ij}$ helps to realise low dimensionality of the Geographic Manifold. And (2) $\Phi_{ij}$ divides the spatial interactions by the "mass", and hence eliminates heterogeneity (Fig. 1). We will provide empirical evidence and formal proof later showing that the local Euclidean and low-dimensional properties are grounded in the nature of spatial interactions, and the evidence for triangular inequality satisfaction as well (see Methods for details). Before these, we utilize two typical spatial interaction problems, location choice and propagation, to examine how the Geographic Manifolds perform.

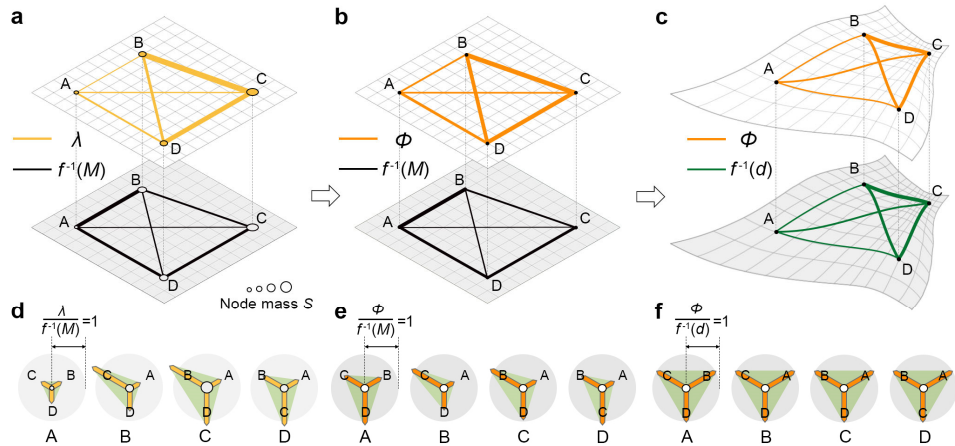

**Fig. 1| Defining the Geographic Manifold. a-c,** three different spatial representations, where the gridlines are the standard grid in spatial representation, which is the reference for geodesic lines, reflecting the geometric difference between different spatial representations. Dots A-D indicate the positions or nodes, and their sizes indicate $S$. Edges between nodes indicate the interaction intensity or distance, where thicker edges in the upper layer indicate stronger $\lambda$ or $\Phi_{ij}$, and thicker edges in the lower layer indicate shorter map distance ($M$) or $d$. **d-f,** the steps for constructing the featureless plain. This is achieved by matching the generalized interaction measures $\lambda$ or $\Phi$ with the respective generalized distance measures $M$ or $d$. The circle centers correspond to nodes A-D, respectively, in panels **a-c**; and the bar size marks the degree of matching, with a perfect match between the two achieved at the radius of the grey circle. **a, d:** Spatial representation on a map. It can be seen that $\lambda$ does not always correspond to $M$; Meanwhile, the overall matching of different nodes in panel **d** varies (e.g., A and C, which can be seen from the size of the light green area in the radar chart), which is caused by the heterogeneity of $S$. **b, e:** Eliminating heterogeneity. Since $\Phi$ by definition normalizes the heterogeneity of $S$ (Eq. (1)), $\Phi$ is much closer to $f^{-1}(M)$ than $\lambda$, and that the average matching across nodes is consistent (i.e., the total size of the bars is consistent) in panel **e**. **c, f:** Geographic Manifolds. By Eq. (1), $d$ eventually achieves a perfect match with $\Phi$, and hence the homogeneous, low-dimensional Geographic Manifold is defined.

## Spatial analytics on Geographic Manifolds

### Location Choice

The location choice problem is one of the most classic spatial analysis problems[13,20,33,34]. The locations of various facilities emerge as a set of controlling nodes in a spatial interaction network, and their controlling nature is reflected in the fact that all the location solutions together provide optimal coverage of the facility needs of a population in motion. Classic location theory, in view of the inability to directly observe the movement of large numbers of people at that time, approximates the movement with distances, and gives a uniform spatial partitioning model[13] under the premise of the featureless plain (Fig. 2a). This simplification becomes unacceptable in practice and theory, especially when recent technological developments have enabled direct observation of mass people movements, which are obviously heterogeneous. An empirical optimal location choice results (see Methods for details) in the real world perfectly illustrate this deviation (Fig. 2b).

On the Geographic Manifold (Fig. 2c; see Supplementary Material 2 for the process of forming the Location Choice Manifold; see Supplementary Material 3 for the Location Choice Manifold with real-world locations), the uniform location solutions are restored (see Methods for details of the homogeneity test). Note that the homogeneous location pattern given by the classic theory actually minimizes the total distance from all locales to the location solutions. If we replace the measure of distance with the repetitive coverage between locales (see Eq. (4) in Methods for details), and provided that the population is uniformly distributed which is easily achieved with cartogram[35,36], the homogeneous location pattern minimizing repetitive coverage is further equivalent to maximizing the total number of people covered by the location solutions. Consequently, a manifold on which distance is defined by repetitive coverage produces again the uniform location solution. As far as human mobility is concerned, the people repetitively covered by two potential location solutions are by definition those who move between them, which is given by $\lambda_{ij} = |c_i \cap c_j|$, where $c_i$ is the set of individuals who visited node $i$. And $\Phi$ will be exactly the repetitive coverage, with $g(S_i, S_j) = \min(S_i, S_j)$ and $S_i = |c_i|$, and operationally, $f = 1/\Phi$ for simplicity. Hence, a manifold defined by $\Phi$ is the spatial representation we need, on which the location choice problem is reduced to a homogeneous partitioning problem.

### Propagation

The propagation problem is a central concern in the study of spatial processes[37], and Hägerstrand developed a classic theory of the problem[21,38]. In the theory, propagation involves a motion-contact process, in which propagation occurs when the intensity of contact exceeds a certain threshold. Then, neglecting heterogeneity, the result of propagation will show a simple pattern of concentric diffusion along the interaction gradient (Fig. 2d)[21]. The theory has broad applicability in the diffusion of political ideas[39], innovation[21], biological subjects[40], and certain types of infectious diseases[10], among others.

However, in real-world situations, such as in an empirical case of a COVID-19 outbreak event in Beijing in June 2022 (see Methods for details), it is clear that heterogeneity of spatial interactions cannot be ignored. The propagation pattern in this case reveals an irregular shape (Fig. 2e, see Methods for details).

To address heterogeneity, we can view monocentric propagation as a distributing process, in which the spatial distribution of infected cases will be determined by the distributing ratios of population movements from the center of propagation to other locales. The distributing ratio here characterizes the motion and contact of propagation conveyors. Note that the distributing ratio happens to correspond with $\Phi$, given that $\lambda_{iO} = n^{\text{prop}}(i; O)$, $g(S_i, S_O) = S_O$, and $S_O = \sum_i n^{\text{prop}}(i; O)$; where O is the propagation center, $n^{\text{prop}}(i; O)$ is the number of infections on $i$ caused by O. Therefore, we can again construct a manifold with $\Phi$. Operationally, $\lambda$ can be characterized by a spatiotemporal co-occurrence matrix, and $f$ takes the logarithmic form according to previous studies[9] (see Methods for details). Applying the approach to the above-mentioned Beijing COVID-19 outbreak event, the results (Fig. 2f) show that the obtained distribution of infected cases on the manifold exhibits a notable concentric pattern (see Methods for the isotropy test).

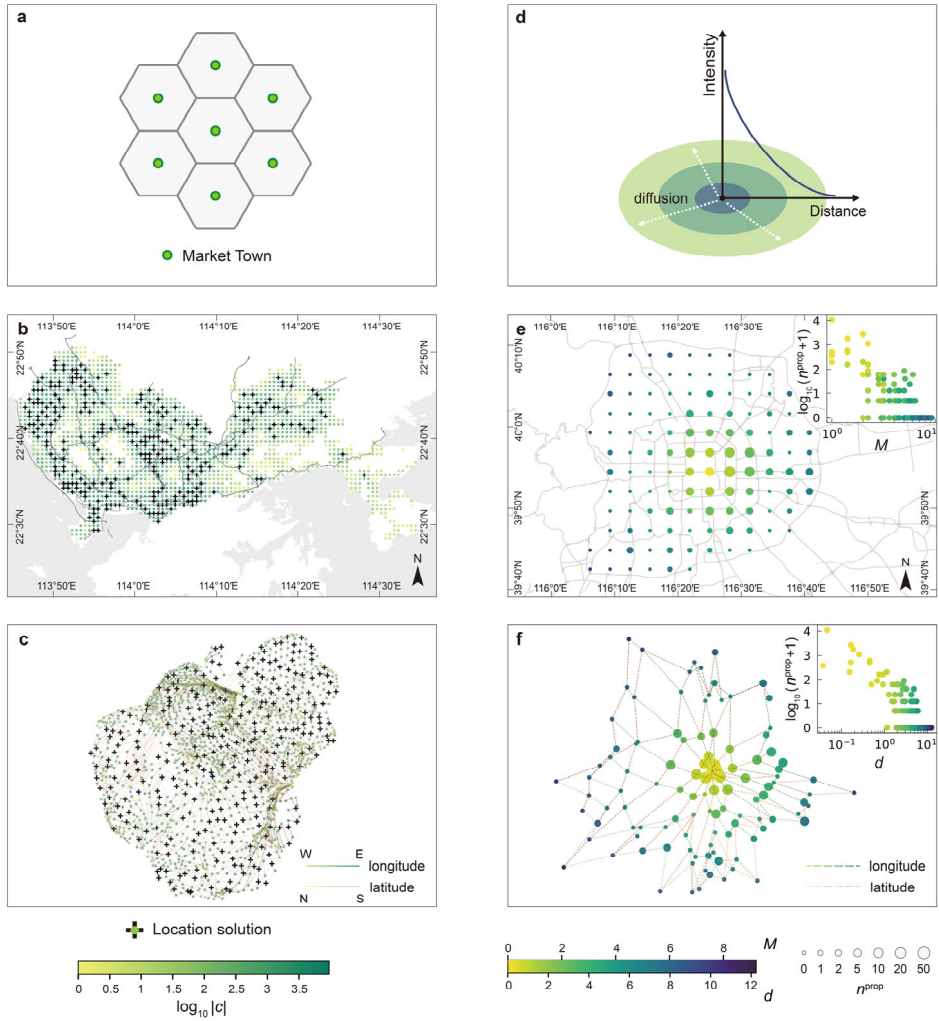

**Fig. 2 | The location choice problem and the propagation problem on a map versus on Geographic Manifolds. a-c,** The location choice problem (the case of Shenzhen). The node (or cell) colors in **b**, **c** depict population in the cell; the cross marks depict location solutions. The classic location choice law under the featureless plain assumption[13] gives rise to regularly laid location patterns **a**, while the simple pattern is broken in real-world location solutions (see Methods) where human mobility and heterogeneity distort the featureless space **b**. Those in the manifold space restores the regular pattern **c**, where the latitude-longitude lines with gradient color trace the deformation of the map on the manifold (with the uniformity index improves from 1.04 in **b** to 2.27 in **c**. See Methods for details of the uniformity test). **d-f,** the propagation problem (the case of Beijing). The node size in **e**, **f** depicts the number of infected cases in that cell; the color of the nodes depicts $M$ and $d$ in the respective spatial representation. The classic propagation law would predict a concentric diffusion pattern of infectious diseases spread in the featureless space **d**. Once again, irregular human mobility in terms of orientations and distances in the real-world rejects the straightforward gradient-

descending pattern with anisotropic dispersions and "jumps" **e**, whereas a proper manifold transformation restores the featureless property, presenting a concentric pattern of propagation **f**, where the meaning of the latitude-longitude lines is the same as they are in **c** (with the $R^2$ of fitting improves from 0.54 in **e** to 0.69 in **f**. See Methods for details of the fitting). For data sources, see Section 1 of the Supplementary Information.

## Existence Proof of Geographic Manifolds

### Empirical Evidence

The two cases above demonstrate the capability of Geographic Manifolds to interpret and present empirical observations in a succinct manner. However, a prerequisite for manifold embedding is that $\Phi$ really exists on a low-dimensional manifold. We will show here that $\Phi$ has local low-dimensional Euclidean properties (the Euclidean properties are discussed in Section 4 of Supplementary Information), providing an empirical basis for Geographic Manifolds.

In our empirical analysis, the local definition is inspired by Isomap's radius approach[41]: centered on each node $i$, all nodes within a certain radius ($d_{\max}$) and all spatial interactions between them form a subgraph, denoted $\mathcal{D}(i; d_{\max})$. The simplices in each dimension, and a generalized clustering coefficient $\theta$ measuring the number of such simplices can be computed for these subgraphs accordingly (see Methods for details). The results show that its low-dimensional subgraph accounts for more than 90% of all instances (Fig. 3 a-c). We demonstrate the robustness of such dimensional analysis of the subgraphs in Sections 4 and 5 of the Supplementary Information.

A natural problem in the above analysis is the impossibility of forming simplices when the neighborhood is sufficiently small. We give an empirical proof for this problem as well: by means of an appropriate annulus expansion method, the concatenation set of a subgraph similar to $\mathcal{D}$ (i.e., the annulus $\mathcal{S}$) can eventually cover the whole manifold space. By the mathematical definition of manifolds, $\mathcal{D}$ gives their local Euclidean properties, and the local expansion, represented by $\mathcal{S}$, will progressively include the intersections between any neighborhoods[27] (Fig. 3d) (See Methods for details). The analysis of $\mathcal{S}$ shows that at all scales[42], spatial interactions are locally embeddable in low-dimensional manifolds.

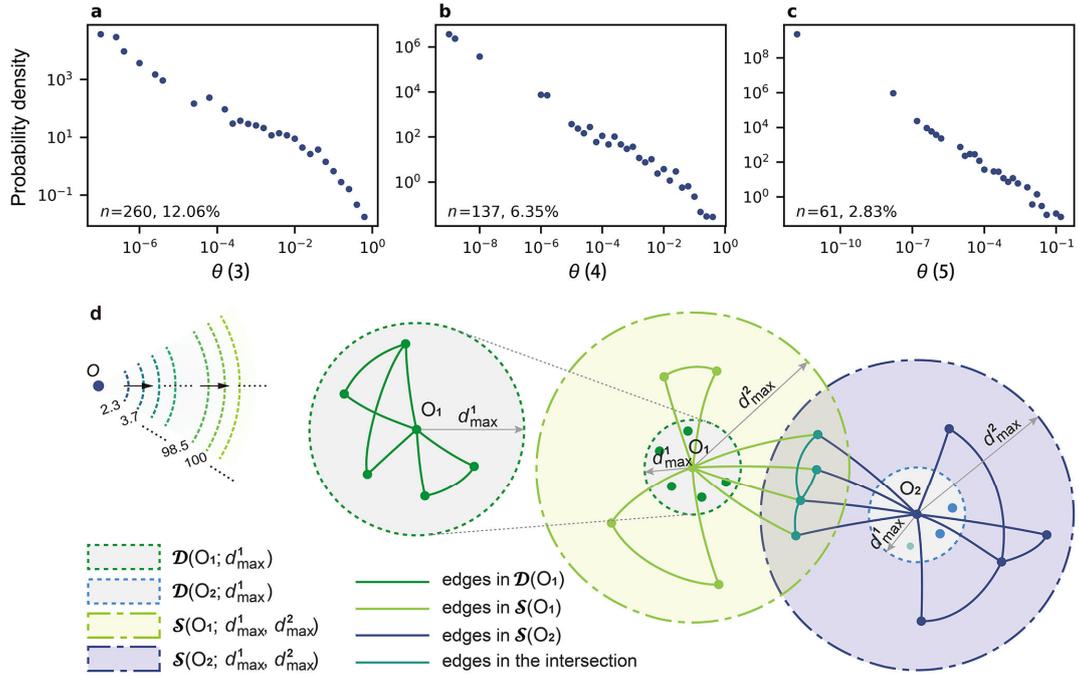

**Fig. 3 | Evidence of the Euclidean low-dimensionality of Geographic Manifolds. a-c,** Clustering coefficient probability distributions of three-, four-, and five-dimensional simplices, respectively, where the notes in the lower-left corner indicate the percentage of neighborhoods with at least one three-, four-, and five-dimensional simplex, respectively (with a total of $N = 2156$ neighborhoods). It can be seen that most of the neighborhoods do not have three- or higher dimensional simplices. $d_{\max}$ is set to 2 in the figure, and results for other $d_{\max}$ are given in Section 4 of the Supplementary Information. **d,** Expanding from local to global space, the spatial interactions remain low-dimensional. Left, empirical results of expanding the radius so that each new expansion of the annulus $\mathcal{S}$ remains low-dimensional, and it can be seen that the maximum radius has exceeded 100, which is inclusive of the whole space for this case. Right, validation of the manifold's low-dimensional property with the local connections within it, where the inner dotted line circles in gray indicate a small $d_{\max}$ neighborhood, and the outer dash-and-dot line circles indicate large $d_{\max}$ neighborhoods. When two large $d_{\max}$ neighborhoods intersect, this intersection will always occur on an annulus with a certain inner and outer diameter. Thus, the local Euclidean dimensional properties of the manifold depend on the Euclidean dimensional properties of $\mathcal{D}$, and the properties of the connection between the localities depend on the annulus $\mathcal{S}$. According to this principle, the empirical proof of the left figure endorses the low-dimensional properties of the whole manifold.

Formal Proof

More rigorously, we proceed to give a mathematical explanation of the existence of Geographic Manifolds, which, as mentioned earlier, is equivalent to the low-dimensional nature of $\Phi$. Given a neighborhood radius $d_{\max}$, the number of high-dimensional neighborhoods can be given by the conditional probability formula:

$$n(K_m \subset \mathcal{D}, \varsigma(\mathcal{D}) = \varsigma | d_{\max}) \tag{2}$$

$$\approx E[P(K_m \subset \mathcal{D} | \varsigma(\mathcal{D}(\cdot; d_{\max})) = \varsigma) \cdot P(\varsigma(\mathcal{D}(\cdot; d_{\max})) = \varsigma) | d_{\max}] \cdot N$$

where the left side of the equation (i.e., the first line of Eq. (2)) is the number of neighborhoods that contain at least one $m$-complete subgraph $K_m$ (and thus the neighborhood is at least $m-1$ dimensional). To the right-hand side of the equation (i.e., the second line of Eq. (2)), $P(K_m \subset \mathcal{D} | \varsigma(\mathcal{D}(\cdot; d_{\max})) = \varsigma)$ is the conditional probability of the occurrence of a simplex in the neighborhood conditional on the number of nodes in the neighborhood (denoted as $\varsigma$), and $P(\varsigma(\mathcal{D}(\cdot; d_{\max})) = \varsigma) | d_{\max})$ is the probability of $\varsigma$.

Considering the characteristics of spatial networks—namely, spatial constraints—we can derive the following theoretical insights for the two terms on the right-hand side of Equation (2). (i) The quantity $\varsigma(\mathcal{D})$ can be considered as a generalized node degree. In spatial networks, the formation of nodes with high degree is subject to spatial constraints such as capacity and distance costs, resulting in a non-power-law distribution of node degrees[14,43]. In other words, spatial networks tend to have fewer neighborhoods with high $\varsigma$ than typical networks. (ii) The probability $P(K_m \subset \mathcal{D} | \varsigma(\mathcal{D}(\cdot; d_{\max})) = \varsigma)$ is influenced by at least two factors: on the one hand, spatial networks usually exhibit disassortativity[14,23], causing the number of edges within $\mathcal{D}$ not large when $\varsigma$ is large; on the other hand, spatial constraints mean that a significant portion of the neighbors of a given node with high $\varsigma$ belong to different, distant regions[42], which are often not tightly connected to each other and do not contribute to the formation of simplices.

Formally, we prove that, first, $\varsigma(\mathcal{D})$ is lognormally distributed based on the rank-size law of $\Phi$ and the Central Limit Theorem (Fig. 4a, b); Second, based on the combinatoric calculation (Fig. 4c), the mathematical expectation of the probability of $\mathcal{D}$ being high-dimensional depends on an asymptotic saturation function (see Methods). The results show that the number of simplex-containing neighborhoods calculated in the empirical evidence is consistent with the theoretical estimation (Fig. 4d, e), and hence the above two mechanisms combined give a reasonable explanation for the existence of Geographic Manifolds. As a conclusion, the empirical existence and theoretical proofs of Geographic Manifolds show that the homogeneous, low-dimensional Geographic Manifold is not merely coincidental; it is a phenomenon that finds grounding in both theoretical and empirical realms, substantiated by the properties of spatial interactions.

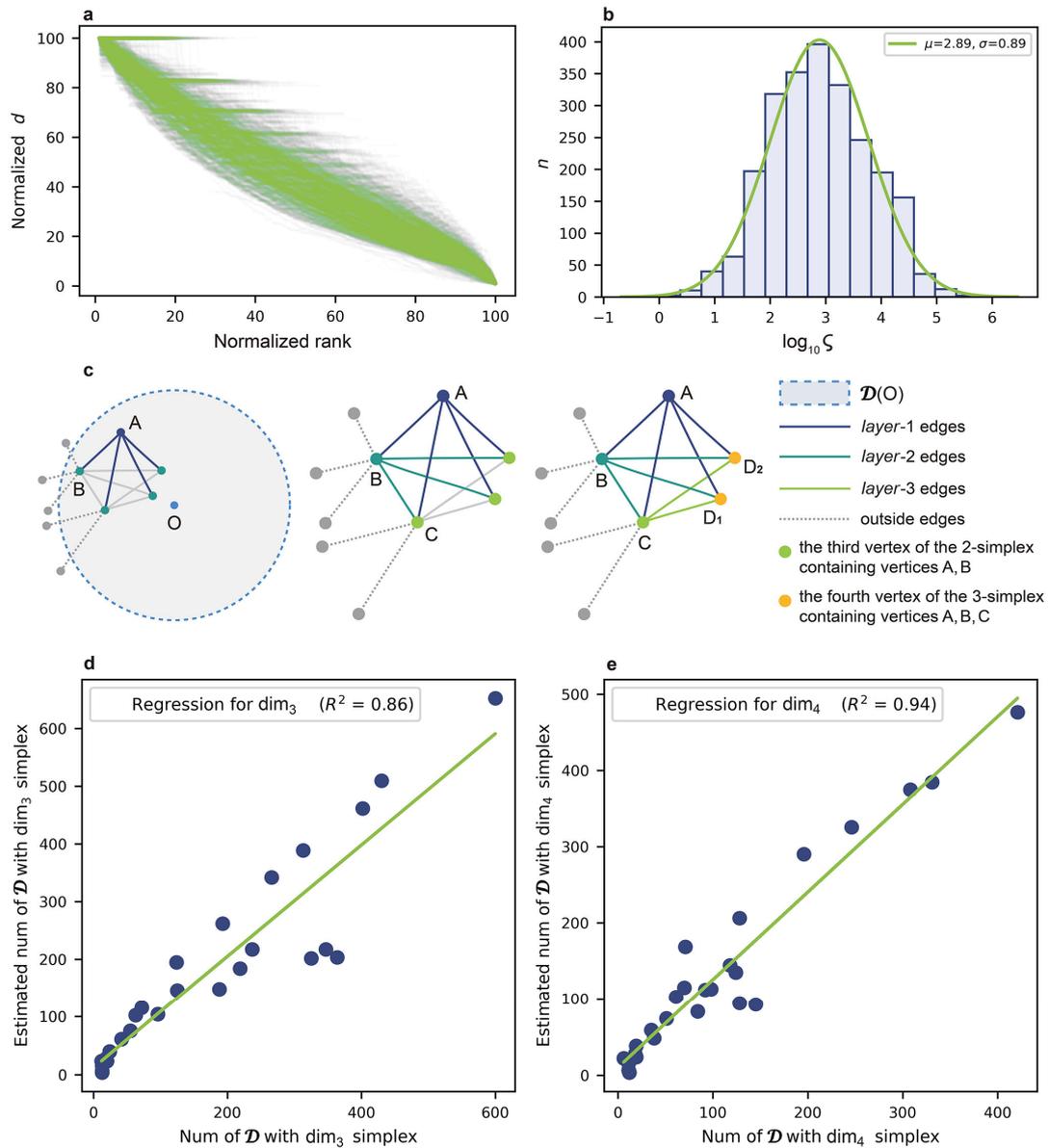

**Fig. 4 | Theoretical explanation of the Euclidean low-dimensionality of Geographic Manifolds. a-b,** First part of the formal proof: we show the log-normal distribution of the number of nodes contained in the neighborhood based on the spatial interaction rank-size law. **a,** the rank-size distribution of $d$, and since $d$ is inverse to $\Phi$, the latter has a rank-size distribution that obeys the power law, which is consistent with the inter-city situation[28] (see Section 4 of the

Supplementary Information for details). **b**, the normal distribution of $\log \varsigma$. See Section 4 of the Supplementary Information for the case under other $d_{\max}$. **c**, Second part of the formal proof: the dimensionality calculation of the local subgraph based on combinatorics. The figure illustrates the case of starting from an arbitrary point A and choosing the appropriate nodes to form the $(m-1)$-dimension simplices, namely $K_m$. If a triangle is to be formed, it is necessary to choose any node B (left) that is connected to A and to find a node C (middle) that is connected to both A and B. If it is to find a tetrahedron that contains A, it is further necessary to find a node D (right) that is connected to both A, B, and C. **d-e**, results from the theoretical calculations of the above two parts for the three- and four-dimensional simplices, respectively against empirical evidence, where the good fitting illustrates the validity of the theoretical explanation.

## Discussion

This paper highlights the significance of $\Phi$-based manifolds as a spatial analytical approach through both empirical and theoretical analyses, specifically by exploring the statistic physics properties of the Gravitational Law (see Section 6 of Supplementary Information for more discussion) and the spatial constraints of spatial networks. The research shows that conclusions drawn from spatial networks and effective distances can be strengthened in adopting a succinct spatial representation. Additionally, the local Euclidean form allows for convenient dialogs with the conventions of classic spatial analytics, while offering a promising simplification of the analysis in the Big Data era with its succinctness. More broadly, aligning with the vision of the founders of modern theoretical geography to reintegrate the millennia-old traditions of geography with geometry[44], the introduction of manifolds in spatial analytics is in essence to regularize geospatial structure in a geometric way. The approach is hence likely to contribute in learning abstract spatial structures for understanding and optimization purposes, where recent advances have already hinted at the effectiveness of a proper spatial representation in spatial analytics for real-world challenges[17,18].

# Methods

## 1. The Location Choice Manifold

Location Choice as a Set Cover Problem

In this study, the optimized location of a set of generic service facilities comes from the set cover problem of maximizing the coverage of the individual clients[45]. Firstly, the individuals' access to each spatial cell (or node) is obtained from trajectory data (thereby aggregating the complete trajectory to the node to protect the privacy of the data provider; see Section 1 of Supplementary Information for the specifics of this data). With the number of chosen locations as the cost, and with the total number of covered individuals as the utility, we can construct the integer programming problem, or its linear approximation as well as a greedy optimization.

$$\max \left| \bigcup_{i \in l} c_i \right|, s.t. |l| \leq \kappa \tag{3}$$

where $c_i$ is the set of individuals covered by node $i$. $l$ is the set of locations, and $l^*$ is the set of optimized locations. $\kappa$ is a predefined parameter that constrains the number of locations to be selected, and $|\cdot|$ denotes the number of elements in the set.

The Location Choice Manifold and Uniformity Test

In the Location Choice Manifold, repetitive coverage is intuitively defined as:

$$\Phi_{ij} = \frac{|c_i \cap c_j|}{\min(|c_i|, |c_j|)} \tag{4}$$

where $|c_i \cap c_j|$ is exactly the OD interaction, thus $\Phi_{ij}$ can be regarded as an extension of the $X$ variable[28] in intra-urban human mobility, and the denominator is taken to be the smaller one of the number of people covered by $i, j$ in order to achieve bi-directional symmetry while ensuring that $\Phi_{ij} \in [0,1]$. The inverse of $\Phi$ is taken as the distance function, and the resulting distance matrix is fed into the t-SNE manifold learning algorithm[46] (see Section 2 of Supplementary Information for an illustration of the choice of t-SNE) to obtain the Location Choice Manifold. The specific form of the distance function, or the manifold learning algorithm does not affect the low-dimensional nature of the Geographic Manifolds, as empirical evidence and formal proof of such low dimensionality is based on $\Phi$ rather than $f(\Phi)$. Subsequently, we transform the Location Choice Manifold with the goal of population homogeneity: First, we perform a Voronoi tessellation of the points on the manifold; then, with the number of individuals covered by the nodes $|c|$, we draw a cartogram[36] of the Voronoi polygons; and finally, we use the geometric centers of the cartogrammed polygons as the nodes in the final Location Choice Manifold (as in Fig. 2c).

The uniformity test of the Location Choice Manifold is as follows[47]: a certain scale of raster is used to define the manifold boundary, and within the manifold boundary, the ratio between the average nearest neighbor distance on the Location Choice Manifold, $d_a$, and the average distance from a Poisson distribution, $d_e$, is computed as a measure of the uniformity of the Location Choice Manifold, $R$. The higher the uniformity index, the more uniform the layout is. The reference value of $R$ and the corresponding uniformity is shown in Section 2 of Supplementary Information. The reason for extracting the manifold boundaries with a raster is that the point cloud on the manifold does not have an overall clumped shape (this is caused by the fact that the data input only retains the nodes through which the trajectory passes, and which are in fact separated by geographic features, such as mountains and rivers on the surface of the earth), thus naïve boundaries such as a convex hull may cause an error. Since the choice of raster scale affects the uniformity results, we choose two different raster scales as sensitivity analyses: the average minimum distance of all nodes, and the average minimum distance of optimal location nodes. The detailed uniformity test results and the sensitivity analyses are shown in Section 2 of Supplementary Information.

It is worth noting that while directly cartogramming the Euclidean map with the goal of homogenizing the population density also improves the uniformity (from 1.04 to 1.97), either the map or cartogrammed map still fall short of the Location Choice Manifold without homogenization of the population (2.20), and much less than the Location Choice Manifold after homogenization of the population (2.27).

## 2. The Propagation Manifold

Data of Infectious Diseases in the Propagation Manifold

For the validation of the Propagation Manifold, we use the spatial distribution of cases from a COVID-19 outbreak in Beijing in June 2022 (commonly referred to as the "Tiantangchaoshi outbreak" because the outbreak started from a bar called Tiantangchaoshi on June 6), with case records from the official website of the Beijing Municipal Health Commission[48]. As of June 22 when the outbreak ended, there were a total of 388 infected cases, of which 361 were first-round cases (i.e., those who had visited Tiantangchaoshi). Considering that Beijing initiated emergency epidemic prevention measures after the first round of epidemic propagation (June 9), which may introduce bias in the modeling, this study only focus on the first-round cases. After geocoding, the residential position of the infected is transformed into latitude and longitude coordinates. There are 349 addresses in the first-round cases that could be used for further analysis, and the remaining 12 addresses are outside the study area (see Section 1 of Supplementary Information for the definition of the study area).

The mobile phone dataset used for the spatial interaction construction of the Propagation Manifold is provided by one of the largest communication operators in China (see Section 1 of Supplementary Information for the introduction of this data[49]).

### The Propagation Manifold

We weightedly fuse the spatiotemporal coexistence interactions (as an expression of contact[50] propagation) with the matrix characterizing the adjacency between nodes (the introduction of the adjacency matrix can be seen as a hybridization of the two mechanisms of skip transmission and surface diffusion[51]; see Section 3 of Supplementary Information for a detailed discussion) to obtain a new spatial interaction, which we call the propagation interaction, denoted as $\lambda_{i,j}^p$, as shown in the following equation:

$$\lambda_{i,j}^p = \chi_{i,j} + b \cdot \delta_{i,j} \tag{5}$$

where $\chi_{i,j}$ is the number of spatiotemporal coexistences between nodes $i,j$. In our monocentric propagation case, one of the nodes in $i,j$ is the cell/node corresponding to Tiantangchaoshi. $\delta_{i,j}$ is the four-way neighboring interaction between nodes $i,j$. $b$ is a parameter that regulates the contribution of neighboring effects to the propagation processes. As stated in the main text, this spatial interaction is proportional to the $\Phi$ interaction since it is defined according to monocentricity. In a multi-round propagation scenario, the form of $\Phi$ will be more pronounced:

$$n^{\text{prop}}(i;\cdot) \propto \sum_j \lambda_{j,O}^p \frac{\lambda_{i,j}^p}{\sum_k \lambda_{k,j}^p} = \sum_j \lambda_{j,O}^p \cdot \Phi_{ij}^{\text{prop}} \tag{6}$$

We use a distance function $f$ with a logarithmic form[9] that transforms propagation interactions into distances:

$$d_{i,j}^{\text{prop}} = f\left(\lambda_{i,j}^p\right) = 1 - \ln\left(\frac{\lambda_{i,j}^p}{\max\left(\lambda_{i,j}^p\right)} + \epsilon\right) \tag{7}$$

where $d_{i,j}^{\text{prop}}$ is the distance to be achieved by the Propagation Manifold, namely an input to the manifold learning. Note that the distance measured by the Propagation Manifold is $\overline{d_{i,j}^{\text{prop}}}$, which may be slightly different from $d_{i,j}^{\text{prop}}$, as the result of fitting residuals. To avoid taking logarithms to 0, we introduce $\epsilon$, which is taken to be $10^{-16}$ in the calculation. In fact, the specific form of this distance function does not affect the concentric pattern of the Propagation Manifold, considering the fact that the interaction quantity inverted by $\widehat{d_{i,j}^{\text{prop}}}$ also obtains a good fit to the contagion. The magnitude of this inverted interaction is the same as the number of spatiotemporal coexistences, which theoretically will be linearly related to the number of infections (see Section 3 of Supplementary Information), omitting the influence of $f$.

We input this distance into the Isomap algorithm[41] to obtain the Propagation Manifold, which is parameterized by the radius of the neighborhood, $r$. The use of the Isomap algorithm and the parameter sensitivity analysis is given and discussed in Section 3 of Supplementary Information.

### Isotropic Test for the Propagation Manifold

We divide the verification of the hypothesis that Propagation Manifold achieve isotropic distance decay into two steps: in the first step, we regress the manifold distances $\widehat{d_{i,O}^{\text{prop}}}$ and infected cases $n^{\text{prop}}(i; O)$ to account for the correctness of the distance decay pattern in general; in the second step we regress in a grouped manner, with groups defined by direction, to account for isotropy.

For the first step, considering that only a very small number of nodes have small enough distances (or strong enough interactions) to the center of the propagation, nodes with a high number of infected cases (see panels in Fig. 2e, f, and Extended Data Fig. 2) are sparse. These facts make distances and the number of cases form skewed distributions. To avoid the result being dominated by these very few nodes that are close to the center and are highly infected in the fitting calculations, we use a double logarithmic linear regression (see Section 3 of Supplementary Information for a discussion of the fitting methodology), as shown in the following equation:

$$log\big(n_{\text{prop}}(i; O) + 1\big) = \alpha \cdot \log\left(\widehat{d_{i,O}^{\text{prop}}}\right) + \beta + \varepsilon \qquad (8)$$

with $R^2 = 0.69$, and the $R^2$ for regression without logarithm (which is a more common practice for contagion analysis[10]) is 0.88.

For the second step, we apply the above regression to each direction group, and compare the 95% confidence intervals for the parameters $\alpha, \beta$. When there is an intersection of these confidence intervals, we consider the Propagation Manifold to be isotropic, or the propagation takes a concentric pattern. The results of this test are shown in Extended Data Fig. 3, which generally pass the isotropy test.

# 3. The Empirical Evidence and the Formal Proof of the Existence of Geographic Manifolds

Triangular Inequality analysis

To verify whether $d$ satisfies the triangular inequality, we first extract all possible combinations of three nodes where the pairwise $\Phi$ are non-zero, i.e., $\{\{A, B, C\} | \Phi_{i,j} > 0, i, j \in \{A, B, C\}\}$ (for those three-node combinations that include 0-$\Phi$ pairs, the 0-$\Phi$ distances can be filled using geodesics to ensure the triangle inequality). For each of these combinations of three nodes, we then sort the lengths of the three edges and denote them as $e_1, e_2, e_3$ in descending order, or $\Phi_1, \Phi_2, \Phi_3$ in ascending order. Next, we select combinations $\{A, B, C\}$ where $\Phi_1 \in (\Phi - \Delta\Phi, \Phi + \Delta\Phi)$. By plotting $e_2$ on the horizontal axis and $e_3$ on the vertical axis, the cases that satisfy the triangular inequality will fall within the region enclosed by the three dashed lines (see Extended Data Fig.1).

Dimensional Computation and Annulus Analysis Based on High Dimensional Simplices

The dimension of a neighborhood $\mathcal{D}$ is determined by the highest dimensional simplex it contains (see Section 4 of Supplementary Information for additional discussion on dimensions and simplices), i.e., $\dim \mathcal{D} + 1 = \max\{m | K_m \subset \mathcal{D}\}$, where $K_m$ is a complete subgraph, or a simplex, containing $m$ nodes. It is worth noting that in a real network, if only a low percentage of nodes form higher dimensional simplices, while the remaining nodes are organized in a low dimensional form, a low dimensional embedding does not cause a significant error. Therefore, the simplex analysis needs to be viewed in the sense of probability distributions. As is the same to the annulus $\mathcal{S}$.

We generalize the clustering coefficients to fit different dimensions, while defining local clustering coefficients in terms of neighborhoods $\mathcal{D}$. Thus, the clustering coefficient of a triangle, or a two-dimensional simplex, is defined as:

$$\theta\big(3; \mathcal{D}(O)\big) = \frac{|K_3|K_3 \subset \mathcal{D}(O)|}{C^3_{\varsigma(\mathcal{D}(O))}} \tag{9}$$

where $\theta(3; \mathcal{D}(O))$ is the generalized clustering coefficient of the neighborhood $\mathcal{D}(O)$

centered at O. The numerator is the number of 3-complete subgraphs contained in the neighborhood, and the denominator is the number of all possible 3-subgraphs. The generalized clustering coefficient for higher dimensions is:

$$\theta\big(m; \mathcal{D}(O)\big) = \frac{|K_m| K_m \subset \mathcal{D}(O)|}{C_{\varsigma(\mathcal{D}(O))}^m} \quad (10)$$

In the definition of a simplex or complete subgraph, we keep all edges with distances less than $d_{\max}$ and delete edges with distances greater than $d_{\max}$. This analysis emphasizes the topology of the spatial interaction, while weakens the influence of distance. In the case where distances are taken into account, the dimension can be determined using the Multidimensional Scaling (MDS) method[52], and the consistency of the two dimension calculations can be found in Section 5 of Supplementary Information. A method of dimensional analysis based on chordless cycles has recently been proposed[26] and a discussion of the relationship of our approach to this method is given in Section 4 of Supplementary Information.

The annulus $\mathcal{S}(i, d_{\max}^1, d_{\max}^2)$ is defined as the subgraph consisting of the nodes that are within $[d_{\max}^1, d_{\max}^2)$ distance of node $i$, plus $i$ itself, and all the connected edges between them. In Fig. 3d in the main text, each round of expansion of $\mathcal{S}$ is made in the following two steps: firstly, the radius of the current $\mathcal{D}$, or the outer diameter of the current $\mathcal{S}$, is used as the inner diameter $d_{\max}^1$ of the new $\mathcal{S}$; secondly, with a learning rate of 0.1, a gradual attempt is made to expand the outer diameter, $d_{\max}^2$, until the proportion of the new annulus's low dimensionality computed by MDS is reduced to 0.8.

The Formal Proof and Explanation of the Existence of Geographic Manifolds

We will give the formal proof and explanation of the low-dimensional nature of $\Phi$-based manifold in this section. The proof uses basic assumptions that are widely recognized in spatial interactions and human mobility[28,53–55].

Central Limit Theorem and the Log-normal Distribution of $\varsigma$

We find that given $d_{\max}$, the $\varsigma$ of each neighborhood $\mathcal{D}$ approximately obeys a log-normal distribution (see Section 4 of Supplementary Information for normal distribution test results). This property can be explained using the Central Limit Theorem. We can start our analysis with the relationship between $\varsigma$ and $\Phi$. In Section 4 of Supplementary Information we empirically show the following equation:

$$\log \Phi = a \cdot \log \text{rank} + b \quad (11)$$

which is consistent with the power law distribution found in[28]. Thus:

$$\varsigma(\mathcal{D}(\cdot; d_{\max})) = n \Leftrightarrow \text{rank}(\Phi = f^{-1}(d_{\max})) = n \quad (12)$$

$$\Leftrightarrow \sum_{\log \text{rank}=0}^{\log n} d \log \Phi(\log \text{rank}) = \log f^{-1}(d_{\max})$$

$$\Leftrightarrow \sum_{\log \Phi=0}^{\log f^{-1}(d_{\max})} d \log \text{rank}(\log \Phi) = \log n$$

where $d \cdot$ is the differential operator, and $d \log \Phi(\log \text{rank})$ measures the change in $\log \Phi$ corresponding to changes in $\log \text{rank}$, while $d \log \text{rank}(\log \Phi)$ is its inverse function. Employing logarithms ensures linearity for both functions, facilitating the assumption that unit variations in $\log \text{rank}$ and $\log \Phi$ follow identical distributions, as empirically demonstrated in Extended Data Fig. 4. This notion of independence stems from the independent and identically distributed $\varepsilon$ noise in the linear regression model (10). Consequently, the $d \log \text{rank}$ across various $\log \Phi$ values emerges as independently and identically distributed random variables. Invoking the Central Limit Theorem, the mean $d \log \text{rank}$ tends toward a normal distribution, i.e. $\sum_{\log \Phi=0}^{\log f^{-1}(d_{\max})} d \log \text{rank}(\log \Phi) / \log f^{-1}(d_{\max}) = \log \varsigma / \log f^{-1}(d_{\max})$ obeys the normal distribution. As $d_{\max}$ is a constant, $\log \varsigma$ has a normal distribution. We give the mean and standard deviation of the $\log \varsigma$ distribution with respect to $d_{\max}$ in Section 4 of Supplementary Information. Thus $P(\log \varsigma(\mathcal{D}(\cdot; d_{\max})) = x)|d_{\max}) = 1/(\sqrt{2\pi} \cdot \sigma_{d_{\max}}) \cdot \exp(-(x - \mu_{d_{\max}})^2 / 2\sigma_{d_{\max}}^2)$.

Simplex Estimation Based on Combinatorics

To estimate the mathematical expectation of the existence of a simplex in a neighborhood $\mathcal{D}(O)$ formally, we first give definitions of some basic concepts. There are two possibilities for the formation of a three-dimensional simplex in a neighborhood: (1) by a triangle that does not contain the central node $O$. By the definition of $\mathcal{D}$, each of these nodes is connected to $O$, so that this triangle plus $O$ is a tetrahedron. And (2) a tetrahedron that does not contain $O$, or that corresponds to a four-dimensional simplex that contains $O$. Simplices of other dimensions also conform to the above formulation. Thus, we only need to compute the number of simplices $\nu_0(\cdot; \mathcal{D})$ that do not contain $O$, and the number of three-dimensional simplices $\nu(4) = \nu_0(3) + \nu_0(4)$.

We use combinatorics to estimate $\nu_0$ and the corresponding expectation of the existence of a certain dimensional simplex. For any node $A$ that is non-$O$ and has a node degree not less than $m - 1$ (no doubt any node in an m-complete subgraph has a node degree at least $m - 1$), we calculate the proportion of nodes with node degree not less than $m-1$ (denoted B) among the neighboring nodes of A (Fig. 4c is the representation of this combinatorics process), denoted as $\rho(A) := |\{B; d_{A,B} \leq d_{\max}, \deg(B) \geq m-1\}|/\deg(A)$, where $|\cdot|$ denotes the number of set elements; the global average of $\rho$ is $\rho(\cdot; \mathcal{L}_1)$. We then calculate the proportion of B's neighboring nodes (denoted as C) which is connected to A and with node degree not less than $m-1$, denoted as $\rho(B; A, \mathcal{L}_2) := |\{C; d_{B,C} \leq d_{\max}, d_{A,C} \leq d_{\max}, \deg(C) \geq m-1\}|/\deg(B)$; the global mean is $\rho(\cdot; \mathcal{L}_2)$). Then the expectation of a triangle containing A without O is equal to the expectation of a C, which is $\deg(A; \mathcal{D}) \cdot \rho(\cdot; \mathcal{L}_1) \cdot \rho(\cdot; \mathcal{L}_2)/3$, where dividing by 3 is for the double counting of the triangle's three vertices. Similarly, the expectation of a tetrahedron containing A without O is $\deg(A; \mathcal{D}) \cdot \rho(\cdot; \mathcal{L}_1) \cdot \rho(\cdot; \mathcal{L}_2) \cdot \rho(\cdot; \mathcal{L}_3)/4$. Thus, the expectation that there exists at least one tetrahedron $K_4$ in $\mathcal{D}$ is $(\rho(\cdot; \mathcal{L}_1) \cdot \rho(\cdot; \mathcal{L}_2)/3 + \rho(\cdot; \mathcal{L}_1) \cdot \rho(\cdot; \mathcal{L}_2) \cdot \rho(\cdot; \mathcal{L}_3)/4) \cdot \sum_{i \neq O} \deg(i; \mathcal{D})$. When this value is greater than 1 we expect that there is at least one simplex of that dimension in $\mathcal{D}$, and $1/(\rho(\cdot; \mathcal{L}_1) \cdot \rho(\cdot; \mathcal{L}_2)/3 + \rho(\cdot; \mathcal{L}_1) \cdot \rho(\cdot; \mathcal{L}_2) \cdot \rho(\cdot; \mathcal{L}_3)/4)$ can be viewed as a threshold $\tau$ on the sum of degrees, for forming a simplex. Similarly, one can derive the expectation that there is at least one $K_m$ in $\mathcal{D}$:

$$E\left(P\left(K_m \subset \mathcal{D} \big| \varsigma(\mathcal{D}(\cdot; d_{\max})) = \varsigma\right)\right) \tag{13}$$

$$= E(P(\sum_{i \in \mathcal{D}, i \neq O} \deg(i; \mathcal{D}) \geq \frac{1}{\frac{\prod_{n=1}^{m-1} \rho(\cdot; \mathcal{L}_n)}{m} + \frac{\prod_{n=1}^{m-2} \rho(\cdot; \mathcal{L}_n)}{m-1}}))$$

$$:= E(P_c(K_m \subset \mathcal{D}))$$

where $E(\cdot)$ denotes the mathematical expectation, and $K_m \subset \mathcal{D}$ denotes that $\mathcal{D}$ contains an $m$-complete subgraph, i.e., the dimension of $\mathcal{D}$ is not less than $m-1$. Therefore, the first row in Eq. (13) is the mathematical expectation of the conditional probability that the dimension of $\mathcal{D}$ is not less than $m-1$, given $d_{\max}, \varsigma$. The second row is the result computed from the sum of the nodes' degrees in $\mathcal{D}$ as deduced from the previous combinatorics method. $P_c(K_m \subset \mathcal{D})$ is the probability of forming a simplex by combinatorics.

The distribution of $\rho(\cdot; \mathcal{L}.)$ at each level is shown in Extended Data Fig. 5. The distribution of $\rho(\cdot; \mathcal{L}.)$ after changing $d_{\max}$ is shown in Section 4 of Supplementary Information. $\rho(\cdot; \mathcal{L}.)$ contains information about the assortativity of spatial networks[23]. Its exact value and mechanism in relation to the hierarchy deserves further exploration, but this is no longer the scope of this paper and we use the empirical average value of $\rho$ for calculation of $\tau$.

The computational result of the above combinatorics depends on the sum of node degrees, i.e., $\sum_{i \neq 0} \deg(i; \mathcal{D})$, and in order to analyze its relationship with the theoretical $\varsigma$, given $d_{\max}$, we compare the sum of node degrees of each neighborhood $\mathcal{D}$ with the $\tau$, to form a set of sequences of 0s or 1s. We sort this sequence by $\varsigma(\mathcal{D})$ and then perform a sliding window average. The relationship between $P_c(K_m \subset \mathcal{D})$ and $\varsigma$, given $d_{\max}$, is shown in Extended Data Fig. 6. Although the $\varsigma$ of each neighborhood is deterministically unchanged given $d_{\max}$, we may consider a conceptual framework where a neighborhood randomly grow to obtain larger $\varsigma$. Thus $\varsigma$ increase is like the addition of a new node contributing to $P_c(K_m \subset \mathcal{D})$. While the imagined growth of $\varsigma$ is stochastic, the increase in this contribution is therefore identically distributed, namely $dE(P_c(K_m \subset \mathcal{D}))$ will grow linearly until $\sum_{i \in \mathcal{D}, i \neq 0} \deg(i; \mathcal{D})$ is stably larger than $\tau$, reaching $P_c(K_m \subset \mathcal{D}) \equiv 1$ asymptotically, as is the linear growth until asymptotic pattern in Extended Data Fig. 6. After numerical regression of the linear growth and asymptotic patterns of $P_c(K_m \subset \mathcal{D})$ with $\varsigma$ for different $d_{\max}$, we can estimate $n(K_m \subset \mathcal{D}, \varsigma(\mathcal{D}) = \varsigma | d_{\max})$, and compare it with the number of high dimensional neighborhoods in the empirical evidence. The results of the above comparison are shown to be highly consistent in the main text Fig. 4**d, e**, with each point representing a sample of neighborhoods with certain $d_{\max}$ and $\varsigma$.

Therefore, the process of theoretical calculation is as follows: (1) for each group of analysis, input $d_{\max}$, $\varsigma$ range, the dimension of the simplex $m$, and the empirical value of $\rho$; (2) in the preparation stage, we compute $\tau$ with respect to $\rho$ and $m$, and use $\tau$ to perform the asymptotic analysis and fitting; (3) we calculate the lognormal distribution parameter of $\varsigma$ with respect to $d_{\max}$; (4) Using the result of (3), we derive the expectation of the total number of neighborhoods within the range of concern; (5) Based on the results of asymptotic fitting, we calculate the proportion of neighborhoods that theoretically contain a $m$-simplex under the given $d_{\max}$ and $\varsigma$. This proportion is then multiplied by the number of neighborhoods given by the $\varsigma$ distribution to obtain the theoretically estimated number of high dimensional neighborhoods; (6) Compare the results of the theoretical calculations of the various groups with the empirical evidence. The fitting performance of (6) implies to what extent the rank-size law[28] of $\Phi$, the assortativity of spatial networks[23], and the asymptotic information can explain the formation of simplices. High consistency of such theoretical analysis with the empirically low dimensional results show the existence of a low dimensional, homogenous Geographic Manifold is theoretically grounded in the nature of spatial interactions.

## 4. Statistics

The regression for the rank-size analysis (n = 2107), the $\sigma, \mu$ of $\varsigma$ depending on $d_{max}$ (n = 2107), the propagation problem (n = 168) and its isotropic test (see Extended Data Fig.3 for ) is implemented using the OLS function of the StatsModel[56] library in Python. The error bars of isotropic test (see Extended Data Fig. 3 for $n$ values) are the 95% CI of the regression parameters. The distribution test of $\log(\varsigma)$ is implemented using the stats.kstest function of the Scipy[57] library in Python.

## Data availability

Raw mobility data are not publicly available to preserve privacy. $\Phi$–level data to reproduce the findings of this study can be requested from the corresponding authors.

## Code availability

The code to replicate this research can be found at jianghezhishi/Geography-as-Manifolds: Code for the paper Geography as Manifolds. (github.com).

## (Refs in Methods / numbering continued)

## Acknowledgements

This work was supported by the National Natural Science Foundation of China (grant number 42371236) and National Key Research and Development Plan of China (grant number 2022YFC3800803). H.J. thanks Guodong Xue for discussions and comments.

## Author contributions

H. Z., L. X., and Y. L. conceptualized the research. H. J., L. X., T. L., and H. Z. designed the research. H. J. and T. L. processed the data, and build the Location Choice Manifold and Propagation Manifold. H. J. did the empirical analysis of manifold dimensions. H. J., J. T, and Z. C. developed the theoretical argument. Z. C. offers advice on mathematical manifold expertise. Y. W. completed the drawing and embellishing of the diagrams. H. J. and L. X. wrote the paper. L. X. and Y. L. supervised the research. All authors discussed the results and reviewed the manuscript.

## Competing interests.

The authors declare no competing interests.

## Correspondence and requests for materials

Correspondence to L. Xu, H. Zhang, and Y. Liu.

**Supplementary Information** is available for this paper.

# Extended Data

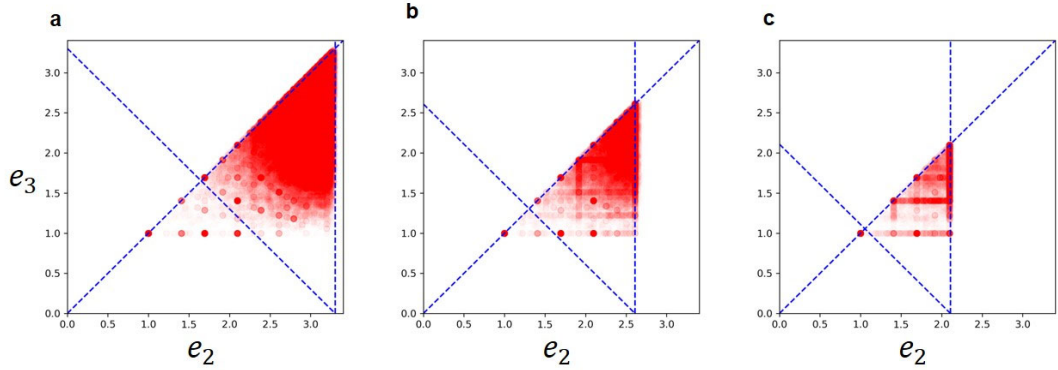

Extended Data Fig.1|Empirical analysis of the triangular inequality. Panels **a-c** show the three node combinations with $\Phi_1 \in (0.09, 0.11), \Phi_1 \in (0.19, 0.21)$, and $\Phi_1 \in (0.32, 0.34)$, respectively, and the distance function takes the form of Eq. (7). The proportion of combinations outside the region enclosed by the three dashed lines in **a-c** are 0.9%, 2.2%, 4.2%, respectively. Therefore, the triangle inequality is statistically valid (for the discussion on the small portion where the triangular inequality does not hold, see Section 6 of Supplementary Information).

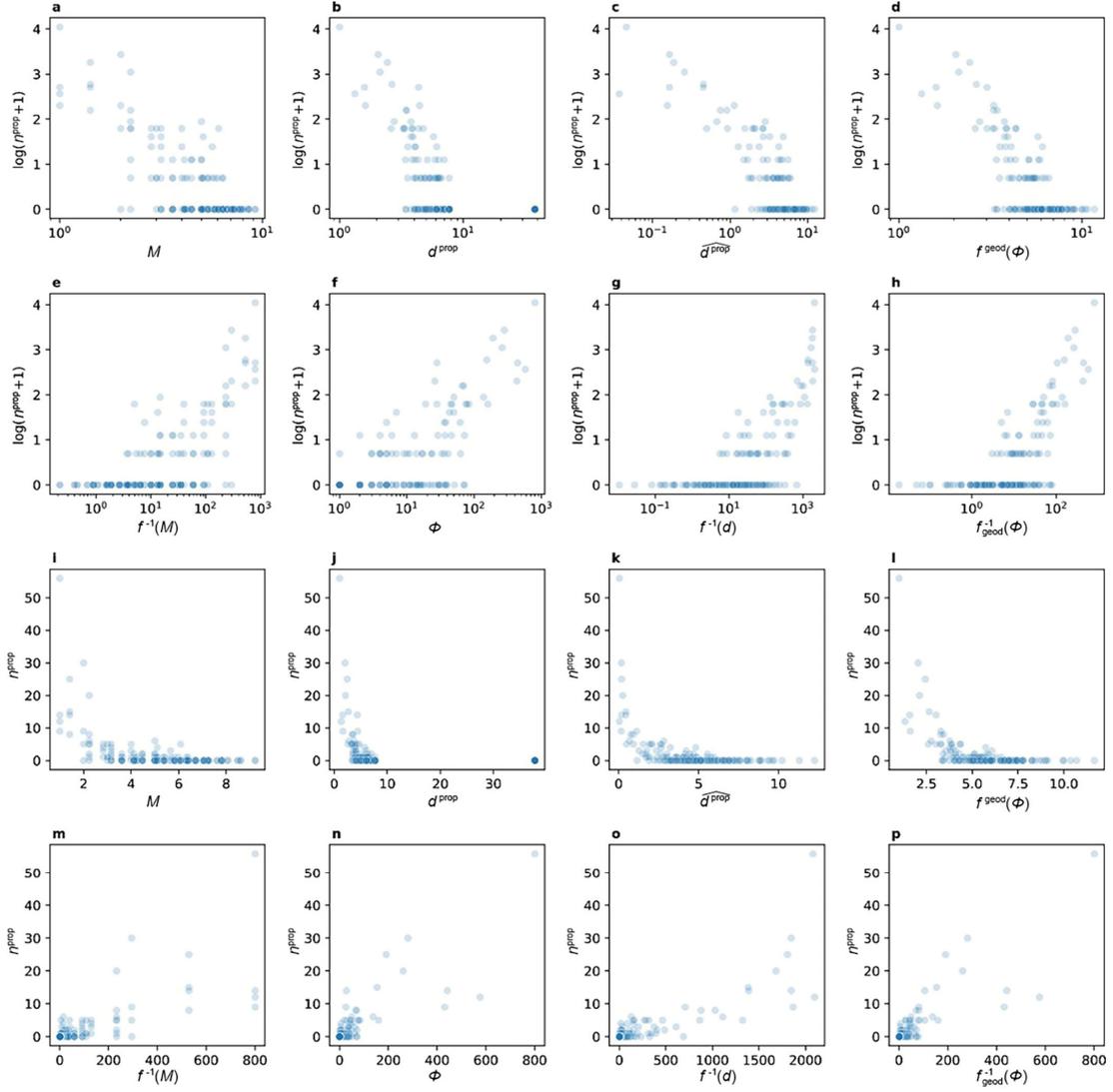

Extended Data Fig.2| Number of infected people versus various distances or interactions. Panels **a-h** show the relationship between the number of infected people and various distances or interactions under the double logarithmic axis, and it can be seen that the scatters of panels **c** and **g** corresponding to the manifold distances or their inverted interactions are more concentrated, forming a clear pattern. Most of the interactions and distances depicted in the figure have been defined in the main text, with the exception of the $f^{\text{geod}}(\Phi)$. The $f^{\text{geod}}$ represents an intermediate state in the Isomap manifold learning process, where distances within the $d_{\text{max}}$ neighborhood are preserved, and distances outside the neighborhood are calculated as geodesic distances. Panels **i-p** show the corresponding results under the non-logarithmic axis, where the linear relationship is still more pronounced in the spatial interaction inverted from manifold distance in panel **o**. This is consistent with the findings that regression $R^2$ of the Propagation Manifold is greater, and thus provides additional supporting evidence to suggest that the Propagation Manifold achieves a better explanation of geographic propagation.

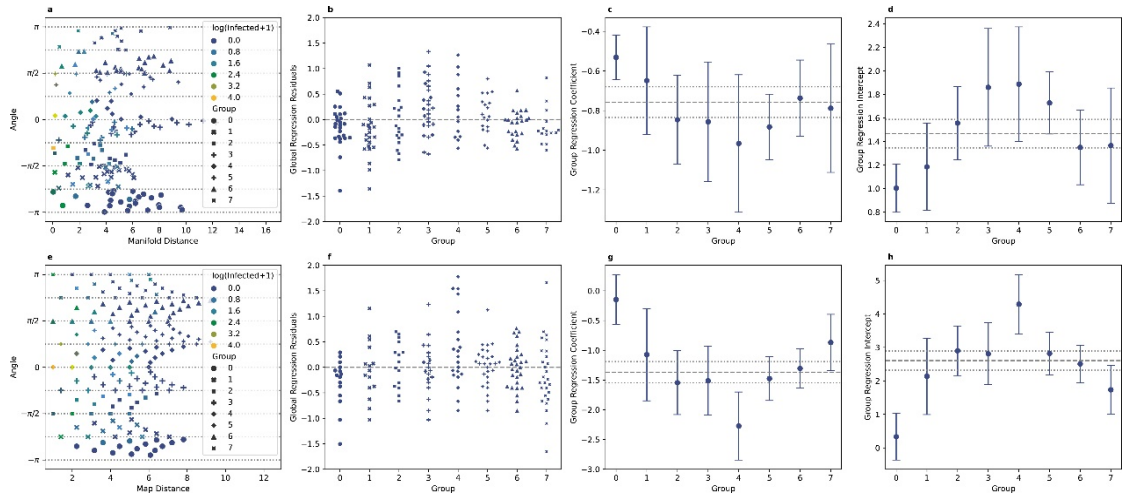

Extended Data Fig.3| Isotropic test results of distance decay on the Propagation Manifold. Panels **a-d** show the isotropic test of the Propagation Manifold ($n = 26, 26, 19, 27, 14, 18, 23, 15$ for group 0-7, respectively). It can be seen that the parameter confidence intervals of almost all directions intersect the global confidence intervals, namely between the two dashed lines ($n = 168$), indicating isotropy. Panels **e-h** show the isotropic test for cases on the map representation ($n = 15, 15, 15, 20, 21, 27, 28, 27$ for group 0-7, respectively), and it can clearly be seen that the isotropy is worse than that on the Propagation Manifold. This phenomenon is evident from Fig. 2**e**, **f** in the main text.

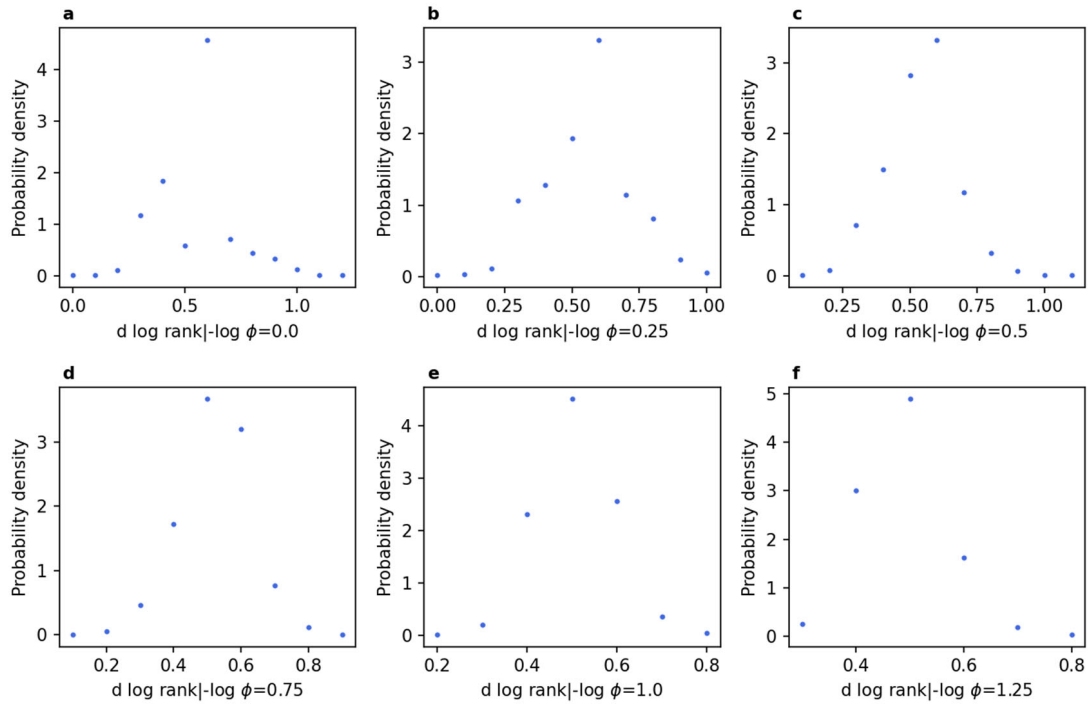

Extended Data Fig.4| The rank order changes caused by the unit $\log\Phi$ changes are approximately identically distributed at different $\Phi$.

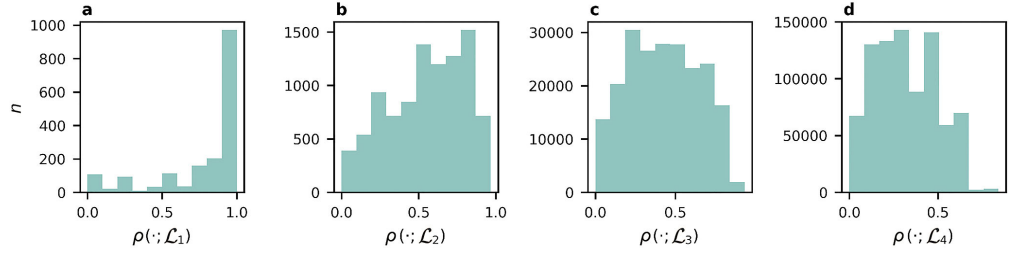

Extended Data Fig.5| Probability distribution of $\rho$ at $d_{\max} = 2$. The average $\rho$ for $layer_1, layer_2, layer_3, layer_4$ are 0.79, 0.55, 0.42, 0.32, respectively. We will give the $\rho$ distributions of other $d_{\max}$ in Section 4 of Supplementary Information.

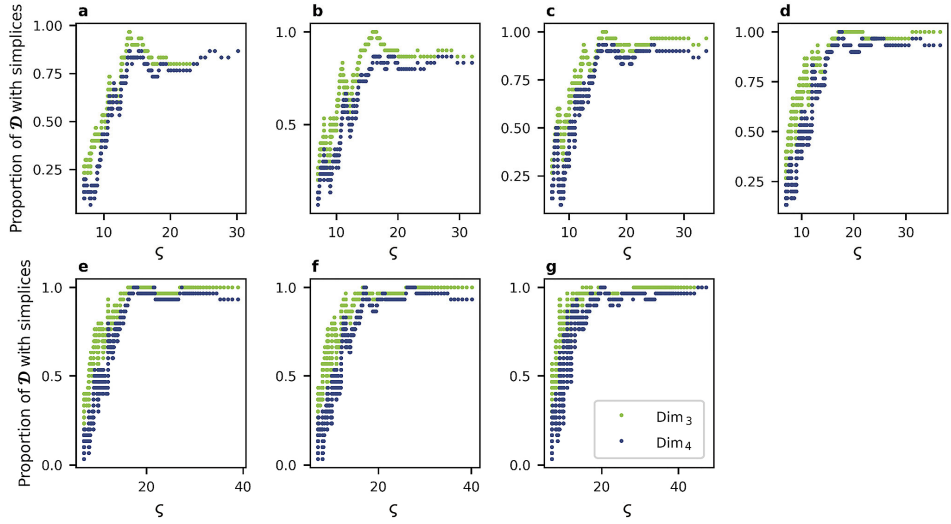

Extended Data Fig.6| Theoretical probability of forming 3-dimensional and 4-dimensional simplices $P_c(K_m \subset \mathcal{D})$ versus $\varsigma$ for different $d_{max}$. Panels **a-g** are for $d_{max} =$ 2, 2.2, 2.4, 2.5, 2.7, 2.8, 3, respectively. From the figures it is clear that a pattern of linear growth followed by asymptotic attainment of 1 is formed.